**Counterfactual Explanations and the Scope of Contestability**

Alice C.W. Huang
Department of Philosophy and Department of Computer Science, University of Western Ontario
Schwartz Reisman Institute for Technology and Society
Email: alice.huang@uwo.ca
Orcid: 0000-0002-1719-1945

Thomas Grote
Department of Computer Science and Artificial Intelligence, University of Technology Nuremberg
Email: thomas.grote@utn.de
Orcid: 0000-0002-9832-6046

*Abstract*: The automation of consequential decisions through opaque machine learning models in societal domains impedes our agency. This paper is about how agency can be reinstated by the provision of certain kinds of knowledge. More precisely, we discuss whether a specific type of explanation, counterfactual explanations, facilitates our ability to contest algorithmic decisions. Against this backdrop, our paper makes three contributions: First, we develop an account of contestability, where contestability is defined as the provision of information, sufficient for a decision-subject to use as a basis for demanding that a decision be revoked. We also demarcate contestability from adjacent concepts in the discourse surrounding the right to explanation, such as justification and recourse. Second, we examine to what extent counterfactual explanations are conducive to contestability by considering a variety of failure modes causing problematic algorithmic decisions and scrutinize to what extent counterfactual explanations help us detect the underlying errors. Third, we propose ways in which, with certain modifications, counterfactual explanations can be made more fitting to serve the desired function. In this vein, we sketch the contours of a multi-shot approach to counterfactuals, where decision-subjects can query a model to test their own counterfactuals for a (limited) number of times.

## 1. *Introduction*

The automation of consequential decisions through opaque machine learning models in societal domains impedes our agency, and concerns about agency often underpin discussions about the right to explanation (Selbst & Barocas, 2018; Jongepier & Keymolen, 2022; Vredenburgh, 2022). This paper is about how agency can be reinstated by the provision of certain kinds of knowledge. More precisely, we discuss whether—and to what extent—a specific type of explanation, counterfactual explanations, facilitates the ability to contest algorithmic decisions, which is central to our agency.

The importance of contestability in the context of algorithmic decisions has been a subject of discussion since the General Data Protection Regulation (GDPR) was enacted. In article 22 of GDPR, it is stated that "the data controller shall implement suitable measures to safeguard the data subject's rights and freedoms and legitimate interests, at least the right to obtain human intervention on the part of the controller, to express his or her point of view and *to contest the decision* (our emphasis)." Subsequently, many have reaffirmed the connection between the ability to contest automated decisions and the right to explanation. Some have argued that the right to explanation is necessary for, and derived from, the right to contest (Mendoza & Bygrave, 2017; Brkan, 2019). Vredenburgh (2022), in her influential paper defending the right to explanation, also regards the ability to contest decisions as being central to informed self-advocacy, which she takes to ground the right to explanation.

While there seems to be a general agreement on the importance of contestability, much less ink has been spilt on what it means exactly for explanations about algorithmic decisions to enable contestability, and the literature is scattered across different subfields (Ploug & Holm, 2020; Kaminski & Urban, 2021; Alfrink et al., 2023; Karusala et al., 2024).

Against this backdrop, our paper has three interrelated goals:

1. We develop an account of contestability, where contestability is defined as the provision of information, sufficient for a decision-subject to use as a basis for demanding that a decision be revoked.
2. We examine to what extent counterfactual explanations are conducive to contestability.
3. We highlight how, with additional procedural support, counterfactual explanations can be made more fitting to serve the desired function.

Counterfactual explanations describe how the output of a machine learning model would change if certain input features were different. These explanations provide insights that are easy to understand by laypeople, for example, by stating that "a loan would have been approved if income were $5,000 higher." To automate the generation of these explanations, often an algorithm is used to search for a modified version of the original input that changes the output while remaining as similar to the original as possible, for a defined metric of similarity.[1]

We focus on counterfactual explanations for two reasons. First, since the publication of the seminal paper by Wachter et al. (2017), counterfactual explanations have become the standard approach in the machine learning literature to help decision-subjects understand algorithmic decisions and navigate changes that could alter future decisions about them. The same authors also stress their importance for contestability, stating that "counterfactuals could be helpful for contesting decisions, and thus provide greater protection for the data subject than currently envisioned by the GDPR……By providing information about the external factors and key variables that contributed to a specific decision, counterfactuals can provide valuable information for data subjects to exercise their right to contest (p. 41)."

[1] The assumption that a similarity metric can sensibly be defined is contentious, and what similarity means could vary depending on the context.

Second, counterfactual explanations are one type of model explanation that appear to sidestep some of the criticisms often directed at advocates of the right to explanation, notably Vredenburgh (2022).[2]

Institutional critiques argue that the right to explanation entails demands for transparency that go way beyond current standards of service industries, public policy, or even best scientific practices. Consequently, they contend that the target of the right to explanation must be profoundly revised or outright refused (Taylor, 2024; Karlan & Kugelberg, 2025). Otherwise, we risk imposing a double-standard between algorithm-based and *traditional* decision-processes, which has undesirable consequences for the design of institutions (see also Zerilli et al., 2019).

Others criticize the adequacy of most types of algorithmic explanations by pointing out that the kinds of explanations that can be generated for algorithmic decisions fall short in their intended goals. Taylor (2024) argues that algorithmic explanations might not be sufficiently sensitive to contextual factors or can fail to capture the real difference makers in an algorithmic decision. Grote & Paulo (2025) claim that the correct interpretation of algorithmic explanations requires a combination of statistical literacy and domain-knowledge that cannot be presupposed for laypeople. They are therefore unlikely to understand the underlying logic of algorithmic decisions to the degree necessary to make real use of that information.[3]

Counterfactual explanations seem to be able to fend off some of these criticisms. For decision-making authorities, generating counterfactual explanations is computationally cheap, model agnostic and side-steps the problem of having to open up the black box (Wachter et al., 2017). This can at least alleviate some of the worries of institutional critiques. Against the inadequacy argument, it has been shown that, given certain conditions, counterfactual explanations are indeed able to capture difference makers (Baron, 2023). Finally, unlike explainability techniques providing summary statistics, they are intuitive—-as psychological research suggests that the capability for past counterfactual reasoning emerges some time between the age of five and twelve in children (for review, Goddu & Gopnik, 2024). Counterfactual explanations therefore seem to be less affected by arguments of intellectual over-demandingness.

Unfortunately, a closer examination of the failure modes leading to faulty algorithmic decisions lends itself to a pessimistic view regarding the extent to which counterfactual explanations provide a sufficient basis for demanding that a decision be revoked. For example, when faulty decisions are a byproduct of errors in data or computation, counterfactual explanations are an inefficient tool for detecting said errors. Moreover, the information necessary to contest decisions is often too complex or might lie outside of what can meaningfully be provided by counterfactual explanations. However, part of our positive contribution is to highlight how some of the problems can be remedied by modifying the way(s) in which counterfactual explanations are currently used. To this end, we sketch the contours of a multi-shot approach to counterfactuals, where decision-subjects can query a model to test their own counterfactuals for a (limited) number of times.

[2] For more thorough reconstructions of the relevant debate, see Dishaw (2025); Grote & Paulo (2025); and Vredenburgh (2024).

[3] There is another variant of criticisms that argue that informed-self advocacy, the interest that Vredenburgh (2022) takes to ground the right to explanation, alone is not sufficient to establish that we have a right to explanation. We either need other more foundational concepts (Grant et al., 2025; Dishaw, 2025), or complementary concepts (Lazar, 2024) to account for why we have a right to explanations.

We proceed as follows: Section 2 develops an account of contestability, using justification and recourse as contrast classes. Section 3 considers a variety of failure modes causing incorrect algorithmic decisions and scrutinizes to what extent counterfactual explanations are able to detect the underlying errors. Section 4 then provides the outlines of a modified approach to counterfactual explanations, better suited to rendering algorithmic decisions contestable.

## *2. Towards an Account of Contestability*

The aim of this section is to develop an account of contestability. To demarcate it from adjacent concepts, we begin by carving up the conceptual space into three distinct kinds of purposes for explanations—justification, recourse and contestability. We make this tripartite distinction as it covers the use of the model as such (justification), in addition to forward-looking (recourse), and backward-looking predictions (contestability).[4]

In setting the stage this way, we take ourselves to be doing something different from, but informed by, the work of those who defend the right to explanation. Our taxonomy here is a high level taxonomy of three broad types of purposes for explanations. We do not take our taxonomy to be exhaustive, but aim to capture many of the purposes that explanations are intended to serve. It is natural that there is a correspondence between reasons for thinking that we have a right to explanation and the purposes one might take explanations to have, even if the mapping is not exact. In the remainder of this section, we first explain what justification and recourse are in our taxonomy as contrast classes before introducing our account of contestability.

### *2.1 Justification*

We use the term "justification" broadly in this high-level taxonomy, encompassing all sorts of considerations that weigh into the question of *whether the decision to deploy or use a model for socially consequential decisions is defensible.* The important difference between justification, on the one hand, and recourse and contestability, on the other hand, is that justification concerns the use of the model as a whole, for the entire population of potential decision-subjects, rather than details about an individual instance of decision.

In this section, we outline some of the considerations often discussed in the literature. For instance, we want to ensure that consequential decisions based on opaque machine learning models are aligned with the rules of our community. We want to have grounds to believe that a decision is fair, that the process follows the rules about how such decisions ought to be made, and that it has been reached on the basis of legitimate considerations. We also want to know that the model is reliable.

[4] Note that our definitions are pragmatic. Since many of the relevant concepts are often used liberally in the literature, our aim is not to find definitions that track all possible uses of the notions.

As a first pass, we can distinguish between *epistemic* and *moral* justification: The former relates to the reliability of the model output, and the latter is concerned with, among others, issues of fairness, legitimacy, or decision-subjects' well-being. Despite being concerned with different kinds of goods, the two types of justification are tightly connected (see also Durán & Jongsma, 2021; Pozzi & Durán, 2024). Epistemic justification might be thought of as a prerequisite for moral justification, since a model that generates its output unreliably will almost inevitably fail to meet demands of fairness or legitimacy (Grant et al., 2025).

If a model is unreliable, then we are not epistemically justified in using it. But establishing reliability is no trivial task. The standard approach in machine learning is to assess a model's predictive performance in practice (the true risk), which we estimate by measuring its loss on a known data set (the empirical risk)—split into training and test data. Given certain conditions, such as the test set being sufficiently large and the probability distribution of the in practice setting being identical to the known data, a model's empirical risk provides a reliable estimate of its true risk (Harman & Kulkarni, 2011; Grote et al., 2024). Crucial, for present purposes, is that reliability is established entirely by inductive considerations on this approach, rendering model opacity epistemically irrelevant (Duede, 2023).

Yet, the practice of machine learning is rife with cases in which a model performs exceedingly well in training, but fails in real world settings—e.g., because the probability distribution has changed over time or because the model exploits confounding information. Some have therefore turned to explainability techniques for detecting such failure cases. However, others have argued that the role of explanations is best understood as a sanity check, and there are more efficient amelioration strategies available, such as better pre-processing of the data or out-of-distribution detection (Freiesleben & Grote, 2023; Baron, 2025).

It is beyond the scope of this paper to provide an exhaustive coverage of the literature on reliability in machine learning. We therefore focus on a few takeaways that are key for distinguishing epistemic justification from contestability. First, since reliability is established through inductive considerations, the locus of evaluation lies on the model's overall *track record*, compared to individual decisions.

Second, evaluating the reliability of machine learning models is not something ordinary decision-subjects can reasonably be expected to perform themselves, since it requires statistical, computational, and methodological expertise. Epistemic justification therefore presupposes a division of labor: Experts are responsible for *producing* the warrant through model documentation, validation, auditing, and post-deployment monitoring, while decision-subjects and the broader public are entitled to *receive* assurance that these expert-generated warrants exist and meet appropriate standards.

Thus understood, the division of labor is analogous to public trust in science (see also Contessa 2023), where laypeople rely on institutional arrangements that ensure competent experts produce reliable results which are then communicated to the broader public (Grote & Paulo, 2025). As we will see, contestability, by contrast, centers on individual decisions, and involves extracting information from the model that is meant to be consumed by laypeople. The contrast between justification and contestability becomes even more pointed when we consider who carries the burden of proof. For contestability, it is the decision-subject who has to detect errors in an algorithmic decision. In comparison, it is the task of developers and decision-making authorities to show that a machine learning model is reliable (Wang et al., 2024).

Compared to epistemic justification, the moral issues at stake for justification can vary greatly, depending on the particular domain and context of application.

For example, the use of opaque machine learning models by public institutions can be at odds with democratic values—raising concerns about whether the relevant institutions exercise their powers legitimately (Lazar, 2024). This means that the institutions should detail the normative rules of the model deployment so that they satisfy public reason requirements. In other words, the "rules, institutions and decisions need to be justifiable by common principles, rather than hinging on controversial propositions which citizens might reasonably reject" (Binns, 2018, p. 545).

On the other hand, if a model is used by a private company to filter applications in a hiring process, we might want to know how the model output is being used, that protected attributes like race or gender have not been considered, that the model performs similarly across different demographics, or that their privacy is being preserved.

Matters are most pressing when it comes to the operationalization of social prediction tasks. It is well-established that social categories are *latent constructs* that are not directly measurable. Model developers therefore have to resort to using observable properties as target variables. The latent construct and the target variable can drift apart, which leads to undesirable downstream effects. Indeed, several cases have already reached notoriety in the relatively short history of fairness in machine learning: Health costs may not adequately capture health needs (Obermeyer et al., 2019); the historical performance of a student's school is not the best predictor for their individual grades (Zimmermann, 2020); and inferring the risk of recidivism from a defendant's criminal history in the past two years punishes members from marginalized and commonly over-policed communities (Angwin et al., 2016). You see the gist. Counteracting potential biases requires construct validation (Jacobs & Wallach 2021; Tal, 2023; Zhao, 2023).

One way to communicate the reliability of machine learning models to the broader public is through model cards, i.e., short documents accompanying (the use) of machine learning models, in which the performance characteristics and value-based decisions in the design and evaluation process are specified (Mitchell et al., 2019; Buijsman, 2024). Just like for epistemic justification, these value-based considerations relate to properties of the model or the socio-technical environment in which the model is being used, as opposed to particular individual decisions.[5]

### *2.2. Recourse*

[5] But see Dishaw (2025) for an opposing view, according to which the justification of an algorithmic decision must convey the reasons for a particular decision.

Recourse is concerned with the forward-looking ability of decision-subjects to navigate the rules of society in such a way that they can achieve their goals (Vredenburgh, 2022). Compared to contestability, there is a more established literature on recourse, with formal definitions, use cases, a technical apparatus, and an explication of the moral underpinnings (Wachter et al., 2017; Karimi et al. 2021; Uphadyay et al., 2021; Venkatasubramanian & Alfano, 2020; Sullivan & Kasirzadeh, 2025).

It is worth clarifying why we characterize recourse as forward-looking and contestability as backward-looking. When we say that recourse is forward-looking, what we mean is that the purpose of recourse is to change a *future* decision, whereas the purpose of contestability is to change a *past* decision. This is true despite the fact that recourse might depend on backward-looking information gathering, such as diagnosing which features were decisive in the prior decision, and whether that decision should be treated as a stable guide for planning. Likewise, contestability might rely on information about the future, such as an approved policy change, to make a case for revoking a decision.

However, recourse and contestability are not simply two sides of the same coin: They bear dissimilarities regarding the nature of the problems that they ought to solve and rely on different (implicit) assumptions.

Consider a paradigmatic case of recourse: Suppose that a machine learning model has been used to filter suitable applicants on the basis of their CVs for a hiring process. Here, recourse means that the rejected applicants are provided with actionable guidance about concrete changes that could lead to a favorable outcome (Karimi et al., 2021). Counterfactual explanations have almost exclusively become the method of choice for recourse recommendations (Wachter et al., 2017). For example, a plausible recourse recommendation could be: *If you had had one additional internship, you would have progressed to the interview stage.* Upon receiving this information, the decision-subject can plan to apply for internship opportunities in order to increase their future chances of a favorable decision.

Two moral philosophical accounts underpin the value of recourse. On the one hand, Venkatasubramanian and Alfano (2020) see recourse as a fundamental good for temporally-extended agency. If we want to be able to successfully plan our lives, we must understand how society is organized in a sufficiently regular, understandable, and corrigible way. By recommending actionable guidance, recourse recommendations lay the grounds for trust that an unfavorable decision can be avoided, so that the decision-subject can non-accidentally accomplish a desired goal—and all the other associated goals that it is a means to.

On the other hand, Sullivan and Verreault-Julien (2022) argue that the capability approach provides a plausible ethical basis for recourse. According to the capability approach, functionings are the various things that a person may value, such as being educated, living a healthy life, having a job, and so on. These functionings are achievements, representing what a person actually manages to be or to do in their lives. Capabilities, on the other hand, are a range of real and substantive options that people can pursue to achieve different functionings. Applied to algorithmic decision-making, the argument is that recourse recommendations provide decision-subjects with options. However, for recourse recommendations to be genuine options conducive to achieving functionings, they must be actionable and sufficiently diverse (i.e., be sensitive to user preferences in terms of the time-frames or associated costs that are necessary to accomplish a desired goal).

The issue of what makes recourse recommendations *actionable* claims center-stage in the technical literature in machine learning. One strategy to flesh out the actionability condition is to impose constraints on the counterfactuals explanations. For example, we might demand that the distance between the factual and counterfactual instances be as close as possible, or that the counterfactuals be sparse in terms of the features involved (Sullivan & Verreault-Julien, 2022). In addition, we could make it a necessary constraint that the counterfactuals in question refer to features that can be meaningfully changed. Returning to the example of the machine learning model used to screen applicants: If a recourse recommendation were to state *if you were three years younger, then you would have gotten an interview*, this would not provide a rejected applicant with an actionable basis.

Yet, even these constraints may not be sufficient for actionability. As Karimi et al. (2021) point out, while counterfactual explanations provide answers to what would need to be different for a desired outcome, they do not specify *how to act* to achieve that outcome. Counterfactual explanations often ignore causal relationships among features—e.g., how income affects savings. As a result, they may require changes that are either infeasible or put unrealistic demands on the decision-subject. The underlying problem here is that directly treating counterfactual explanations as actionable assumes independence between features, which rarely ever holds true in real-world settings. Therefore, the authors advocate for a shift from recourse via the nearest counterfactuals to recourse via minimal interventions. These interventions are modeled by way of structural causal models, capturing both direct changes and downstream effects. Ultimately, the goal is to identify minimal cost-feasible actions leading to a favorable outcome by the model.[6]

To set the stage for a comparison with contestability in the next sub-section, it is imperative to look at some (implicit) assumptions that prevail in the literature on recourse. Recourse recommendations presuppose a context where the decision-subjects are laypeople, and the focus is on individual decisions. It also presumes a *cooperative* setting, where the decision-subject takes it for granted that the decision by the model is unproblematic, and that the recourse recommendation is faithful to the internal logic of the model.

Finally, it is also important to note that recourse recommendations themselves can be conceived as interventions that can be used by model authorities to *steer* the behavior of decision-subjects (see also Perdomo Silva, 2023). For instance, they can incentivize decision-subjects to make changes regarding their financial situation or educational choices; and while these changes may seem to be in their best interests, some might worry that such interventions are paternalistic in a problematic way (Sullivan & Kasirzadeh, 2025).

[6] For an accessible treatment concerning the relationship between counterfactual explanations and causal interventions, see also Buijsman (2022) and Baron (2023).

Having laid down the basic assumptions of justification and recourse, we now turn to defining contestability.

*2.3 Contestability*

Although it is widely agreed upon that contestability is an important goal of explainability (Wachter et al., 2017; Venkatasubramanian & Alfano, 2020; Selbst & Barocas, 2018; Kaminski & Urban, 2021; Vredenburgh, 2022), few have spelled out what it means exactly for explanations to enable us to contest an algorithmic decision. More conceptual groundwork needs to be done. In what follows, we propose and motivate a definition of contestability. To this end, we will draw from the previously developed accounts of justification and recourse, using them as contrast classes.

We define contestability as the provision of information that a decision-subject can use as a basis for demanding that a decision be revoked. An important characteristic of this definition is that it is *narrow* in scope. Contestability is relevant when an individual decision-subject receives an unfavorable decision, and just like for recourse, we assume that the decision-subject is a layperson. However, as contestability is about altering a *past* decision, it is a backward-looking ability.

For simplicity, we assume that the decision is fully automated, in order to rule out cases where the model merely provides input to a human authority, who makes the final decision. Note, however, that this definition permits that the decision-subject is assisted by experts, helping them to interpret and scrutinize the relevant information, as suggested in Vredenburgh (2022). The assumption of full automation is unrealistic in some cases, in particular because in many real-world domains, there are formal requirements for human-in-the-loop.[7] In the case where humans have the final say, and the machine decisions are merely treated as assistance or recommendations to the human decision maker, the decision-subject's ability to contest is further complicated by the need for understanding how the human made the decision in light of the model recommendations. This is because even if the model predictions are good, the human might nevertheless use that information in ways that could be subjected to contestation. The same goes in the other direction. Even if the model recommendations are shown to be problematic, the human might be able to make up for these problems by exercising their discretion. This is a complicated issue whose full treatment deserves a whole paper on its own, so we set it aside and focus first on the simpler case of full automation. 'Revocation' is meant as a shorthand for a broader family of remedial actions, including a demand for review, correction, reconsideration, or a rerun of the decision procedure.[8]

Our account of contestability is much narrower than that proposed by Alfrink et al. (2023), on which "contestability helps to protect against fallible, unaccountable, illegitimate, and unjust algorithmic decision-making, by ensuring the possibility of human intervention as part of a procedural relationship between decision-subjects and human controllers" (p. 616). On their definition, many of the concerns that we discussed under *justification for deployment* are part of contestability. We prefer our narrower notion because we think that there is value in separating the two types of goals if part of the purpose of this conceptual work is to guide which types of explanations or safeguards we should develop. In particular, the kinds of information required for justifying the deployment of a model for a whole population could very much be different from the information required for decision-subjects to understand individual decisions about them and make use of this information.

[7] We thank an anonymous reviewer for pointing this out.
[8] We owe this formulation to an anonymous reviewer.

We speak of the provision of 'information', which is a less restrictive epistemic notion than 'explanation'. Information about how uncertain the model is with regards to a particular decision, for example, is not really an explanation, but can nevertheless be helpful in making a case that an algorithmic decision should be overturned. Moreover, we do not assume that the information must be about the *internal logic* of the machine learning model, for two reasons. First, it is a particular appeal of certain explainability techniques, most pertinently counterfactual explanations, that they can be generated without opening the black box (Wachter et al., 2017). Second, as we will argue in the next section, in some cases where machine learning models make erroneous decisions, the source of error lies beyond the internal logic of the model.

It is important that the information provided counts as compelling evidence, so that a decision-subject has an actionable claim for demanding that a decision be revoked, instead of just providing *any* evidence that an unfavorable decision could be erroneous. However, since a decision-subject's ability to meaningfully contest a decision also hinges on certain background conditions—-e.g., the existence of appropriate legal guardrails—we use the qualifier 'as a basis'. What counts as 'compelling' is context-dependent, i.e., it is determined by the standards of evidence in a social practice or legal framework. In the context of contestability, we can distinguish between (at least) two standards: (i) A triggering standard, where the information is sufficient to initiate review or further investigation; and (ii) an overturn standard, where the information is sufficient for the original decision to be overturned. In light of our broader definition of 'contestability', we adopt the former, less stringent standard: The evidence is compelling if it is enough to show that there is a serious possibility of error, such that a review procedure should be triggered.[9]

In contrast to recourse, in the context of contestability, the relationship between the model authorities and the decision-subject is *adversarial*: The initial assumption of the decision-subject is that an unfavorable decision is based on an error, or otherwise problematic. Moreover, it might not be in the best interest of an institution or a company that the decision-subject detects the error underlying an algorithmic decision—since this might culminate in a loss of trust or even liability charges. In some cases, model authorities might therefore be incentivized to mask potential errors in the model output (Sullivan & Kasirzadeh, 2025). Another distinction from recourse is that, depending on the specific failure mode, the information required to support a demand that the decision be revoked can be either simple or highly complex.

The objective of this section was to establish the conceptual groundwork. In the next section, we examine whether and to what extent counterfactual explanations facilitate contestability.

## *3. From Counterfactual Explanations to Contestability?*

[9] We owe this formulation to an anonymous reviewer.

The right to explanation has faced criticism due to high costs and the difficulty for laypeople to understand (Taylor, 2024; Grote & Paulo, 2025). Counterfactual explanations arguably offer a way to bypass these difficulties, since they are relatively cheap to generate computationally, and it is easy for a layperson to understand what the counterfactual says. That said, much less has been said about whether and how counterfactual explanations support contestability. Applying the definition of contestability we put forth in the previous section, the question we take up in this section is this: Do counterfactual explanations provide information that a decision-subject can use as a basis for demanding that a decision be revoked? If so, how?

To begin answering this question, we walk through different possible bases for demanding that an algorithmic decision be revoked. This list is by no means exhaustive—one cannot possibly imagine every possible scenario where a decision should be revoked. Still, the hope is to outline a taxonomy that covers the most common cases.

*3.1 Case 1: Discrimination*

One of the most frequently discussed grounds for contesting a decision in the literature is discrimination. One way to establish discrimination is demonstrating that a decision-subject received a negative decision in virtue of their membership in a particular social group. For example, in the case of Kyle Behm, discussed in Vredenburgh (2022), a discrimination lawsuit was filed on the ground that the algorithm considers results of personality tests, which are irrelevant to his performance in low-skill service jobs, and lead to candidates being rejected in virtue of having certain mental health conditions (in Behm's case, bipolar disorder).

In Behm's case, there is one counterfactual that is particularly relevant: *If he had not scored badly on personality tests, he would have been offered an interview for those jobs.* If this counterfactual is not true of the model, then it is difficult to see how their case could be built, since it cannot be said that Behm was rejected because of his results in personality tests.

On the other hand, if the counterfactual is true, this information can presumably be used as a basis to demand that the decisions be revoked, so this particular counterfactual explanation supports contestability on our account.

In Behm's case, however, the counterfactual explanation alone is insufficient because it remains to be established that he scored badly in the personality test due to his mental illness, and that this mental illness is irrelevant (or insufficiently relevant) to his job performance. In some other cases, the permuted feature might be more straightforwardly prohibited in the given context—e.g., If you had been white, you would have been offered an interview—and might be alone sufficient as a basis to contest (Kasirzadeh & Smart, 2021).[10]

Two things are worth noting. First, on our account, contestability concerns individual decisions, unlike justification, which concerns the use of the model. Even if a model performs similarly across different demographic groups, an individual might still claim that it discriminates against them. The opposite is also true. A model might unjustly incur harm on a group as a whole, but not necessarily every individual in that group.

[10] The issue of when *all else being equal* changes to a protected attribute like race, age, or disability causes discrimination is technically, metaphysically, and legally complex. Aside from instances of *direct* discrimination, the discrimination can also happen *indirectly* by the model exploiting proxy features (a well-known example is ZIP codes being a proxy for race in highly segregated areas). Hu (2023) discusses moral and metaphysical questions of proxy discrimination. In addition, there is the issue of *intersectional discrimination*, where the interplay of various identity factors creates forms of discrimination sui generis. The locus classicus here is Crenshaw (1989).

Our claim is that a counterfactual explanation—such as *if you had been white, you would have been offered an interview*—can provide grounds for a decision-subject to demand revocation of *their* decision. However, the fact that one individual received a negative outcome that would have been positive if they were not a member of group G does not directly imply discrimination against group G as a whole. The claim one can make based on a counterfactual explanation is different from what group-based statistical criteria try to do, which is to quantify whether a model distributes resources or harm in an unjust way. While one problematic counterfactual may suggest the existence of others, our argument here is limited to the individual case.

Second, counterfactual explanations are subject to the 'Rashomon problem': There is not one unique counterfactual explanation for each decision (Molnar, 2019; Breiman, 2001). It might be true that if Behm had scored higher on his personality test, he would have been hired. It might also be true that if he had had more work experience, he would have been hired. Moreover, if he had scored higher on his personality test, and at the same time he had had more work experience, he would have also been hired. You get the idea. There are many ways to permute the input data to turn a negative decision into a positive one. For this very reason, in the literature on recourse, much of the technical and philosophical discussions focus on what sorts of counterfactual explanations should be provided to a decision-subject: Is it the one that requires minimal cost to change? Is it the dimension along which a data point is the closest to the decision boundary? Should we provide multiple options so that the decision-subject could simply choose which path to take (Wachter et al., 2017; Karimi et al. 2021; Sullivan & Verreault-Julien, 2022)?

The fact that multiple counterfactuals can be simultaneously true complicates the idea that counterfactual explanations can ground a demand for revocation. Counterfactual explanations are often generated by searching for the "smallest" change, defined relative to a context-dependent and often contentious similarity metric. A central issue is whether a decision subject may challenge the outcome only on the basis of this minimal counterfactual, or whether they may rely on any counterfactual that would have flipped the decision. We assume the latter. For instance, suppose the model indicates that Kenji would have been hired if he had one additional year of experience, and that this is the "minimal change." Suppose it is also true that he would have been hired if he were white, even though this is not minimal according to the metric. We argue that he should still be entitled to use that non-minimal counterfactual as grounds for demanding revocation. This is because the assumption that there can be a sensible definition of 'minimal change' or 'most similar input features' is questionable, since similarity can be highly context dependent and difficult to quantify. Without requiring the counterfactual explanation to be minimal, we can say as a general rule that a non-minimal counterfactual is contestation-relevant when the varied feature is inadmissible under the relevant normative, institutional, or legal framework, and when the counterfactual dependence is robust enough to trigger review. This is also why examining multiple counterfactual explanations is often necessary to fully assess whether a decision is illegitimate or erroneous.

The Rashomon problem is particularly tricky once contestability enters the picture. Notice, importantly, that the counterfactuals useful for contestability are very different from those useful for recourse. For the purpose of contestability, there is a set of potentially inadmissible features that need to be permuted, and these tend to be disjoint from the features relevant for recourse. This is because inadmissible features tend to be those that the decision-subject cannot (easily) change at will, such as race, gender, age etc., whereas only features that can be altered are relevant for recourse. This means that we cannot hope that a single counterfactual explanation will serve both purposes.

A counterfactual explanation showing that altering only an inadmissible feature would have changed the outcome provides evidence that the decision may be morally and legally problematic. It exposes a moral error by demonstrating that the model treated the individual not as an equal moral agent but as someone whose opportunities depended on characteristics that have no legitimate connection to merit or qualification, given that all other features were held constant. It may also indicate potential legal violations when the implicated attribute falls within the legally defined list of protected attributes that decision-makers are prohibited from using in decision making. For self-advocacy, these counterfactuals are especially powerful because they provide concrete, intelligible evidence that a person can understand and act upon to get a past decision reversed, without having to make changes in their lives and go through a new decision process in the future.

The Rashomon problem is significant because the roles of recourse and contestability are complementary and, importantly, not mutually replaceable. We follow Sullivan and Verreault-Julien (2022) in understanding recourse to be demanding a kind of *recommendation*, as opposed to reason-giving explanation or evidence (even though the counterfactuals used in recourse are usually still called "explanations" in the literature).[11] Recourse alone is insufficient to address whether the decision itself was fair or permissible, and does not help uncover or correct underlying moral or legal errors that a decision system may be committing. Moreover, there is an underlying assumption behind recourse that it is up to the decision-subject to make changes in order to improve the outcome, which is misleading when the decision is itself morally problematic (Sullivan & Kasirzadeh, 2025).

There are some further difficulties with using counterfactual explanations for contestability. First, if there are multiple inadmissible features, the cost of providing the set of counterfactual explanations necessary to enable contestability increases drastically. Consider the three following counterfactuals:

(a) If you were white, you would have been offered an interview.
(b) If you were a man, you would have been offered an interview.
(c) If you were a white man, you would have been offered an interview.

[11] There is an important further question about what kinds of reason are given in these explanations, and what kinds of reason ought to be given for the different purposes such as contestability and justification (Babic & Cohen, 2023; Hadfield, 2021; Baum et al., 2022; Glock & Schmidt, 2021). While this is no doubt a deeply interesting and crucial question, we set it aside in this paper.

It is possible that in a model, (a) and (b) are false, but (c) is true. But even if only (c) is true, (c) alone provides information that can be used as a basis to demand that a decision be revoked on the ground of discrimination. So to really rule out discrimination, we need to check them all. But this introduces cost-related trouble for counterfactual-based informed advocacy. Although a single counterfactual explanation is relatively cheap to generate, the number of counterfactuals we need to provide to really enable contestability grows exponentially with the number of potentially inadmissible features.

There is a further problem that is analogous to finding the right recourse explanation given the decision-subject's personal circumstances. In Behm's case, he had previous job reports that indicate that his bipolar disorder did not affect his performance, but another candidate who also scored low in the personality test might not be able to establish the same thing. The past job performances of this other candidate might have been affected by their poor mental health, unlike Behm. An identical counterfactual explanation, when applied to different individuals, might enable contestability for one but not the other, so it is not straightforward to know which counterfactuals need to be included in the contestability package.

Since contestability, unlike recourse, is adversarial, if the decision-subject is not provided with the full range of counterfactuals, they risk being vulnerable to what Sullivan and Kasirzadeh (2025) call 'explanation hacking', where decision makers can pick and choose from the set of counterfactuals to only provide explanations that serve their own interests, rather than the decision-subject's. Sullivan and Kasirzadeh's solution to explanation hacking is that we should provide the explanation that best allows the decision-subject to understand *why* the decision was made. But it is not clear that this is satisfactory for the purpose of contestability. Even if the *primary reason,* however technically defined, for someone's rejection is not their race, the mere fact that *if they had been a different race, they would have been accepted, holding all else fixed* should be enough for them to contest. Offering the primary reason of the decision to help the subject understand why the decision about them is made remains too limited.

The takeaway, then, is that while the *right* counterfactual explanations provide a useful basis for demanding that a decision be revoked, it is far from clear how we can ensure that the decision subject receives those particular counterfactual explanations that they need to contest.

### *3.2 Case 2: Failure to Generalize*

Even if there is no problematic dependence of the decision on inadmissible features, the decision-subject might still be able to contest on the ground that patterns learned by the model cannot be generalized to their particular case. Note that here we are assuming that before deployment, the model has been properly tested and its general accuracy has been established. Thus, the complaint is not that the model is inaccurate, but that its prediction is likely incorrect for this specific person.

Suppose a hospital launches a model that predicts risk of chronic fatigue, trained on lifestyle choices and basic health history. Yaojun, a vegan, is predicted to be at a high risk of chronic fatigue. Suppose further that the counterfactual explanation provided says: *If you were not vegan, you would not have been predicted to be of high risk of chronic fatigue*. Provided that it is well known that vegan diets tend to contain less iron, which in turn often lead to chronic fatigue due to iron-deficiency anemia, we might think veganism is an admissible predictor in this example.

Still, Yaojun might contest on the ground that the pattern learned by the model generalizes poorly to his particular case. The trend learned by the model does not apply to Yaojun because he has been diagnosed with thalassemia. Instead of insufficient iron intake, patients with thalassemia have the opposite problem. They tend to have too much iron, and therefore need to make sure that they do not have too much iron intake. Therefore, in Yaojun's case, the causal chain inferred by the model does not apply.

In this case, Yaojun could use the counterfactual explanation to argue that, while the model may be accurate overall, the local patterns it relies on for his data point do not generalize to his situation. What he needs is a reason why, despite the model's tested reliability, it is likely wrong in his case. In other words, if the model is 95% accurate, Yaojun needs to argue that he likely belongs to the 5% that the model is mistaken about. Counterfactuals expose local patterns in the subject's data, and if those patterns are flawed, that alone is evidence he may fall within the small percentage of cases where the model errs.

Unfortunately, the Rashomon problem remains a difficult one. To know which counterfactual explanation would be the most useful for contestability based on this line of argument amounts to knowing where the weaknesses of the model are, and why the model makes those mistakes. This is often a tall ask. (If it were easy, we would have tried to fix the mistakes already.) Automating the generation of counterfactual explanations, then, will be difficult, and likely very costly.

Moreover, realizing that a counterfactual is evidence that the decision-subject likely falls into the 5% of mistakes made by the model requires enough domain knowledge. In this example, Yaojun needs to have background knowledge about the nutritional profile of vegan diets, the connection between dietary iron deficiency and anemia, that chronic fatigue is a symptom of anemia, as well as knowledge about thalassemia. This, again, is a tall ask. Here, counterfactual explanations do not appear to resolve concerns about cost and the need for advanced statistical and domain literacy.

### *3.3 Case 3: Other Errors*

In other cases, a decision-subject has a legitimate claim for a decision to be revoked, but counterfactual explanations are inefficient or entirely unhelpful. For example, suppose Aisha is trying to qualify for a mortgage to buy her first house. One of the factors that determine a candidate's eligibility is debt-to-income ratio. The bank determines eligibility using an automated system, and Aisha's application was denied because her debt-to-income ratio is too high. This might be the result of a clerical error, incorrect formula for computing the ratio, or because the data recorded were outdated and no longer reflect her current financial situation. In this case, Aisha has a legitimate claim for the decision to be revoked.[12]

But how can Aisha know that her denial was due to an error in the input data? Explanations that come in the form of counterfactuals do not efficiently provide her with relevant information. The relevant counterfactual explanation that will reveal to Aisha that there has been an error in her record is presumably something like this: *If your debt-to-income ratio had been lower than $x$, you would have been approved*. Upon seeing this, Aisha will know that there has been an error, since she can calculate that her debt-to-income ratio is already below $x$. But because errors like these tend to be random, there is no way to ensure that the counterfactual explanation that Aisha is provided with is the one that will reveal the error.

[12] Wachter et al. (2017) discuss cases similar to this.

A much easier and better way to ensure there is no input error is to directly confirm with the decision-subjects that the record about them that the bank has on file is correct and up to date. This guarantees that any error will be identified, allowing the bank to rerun the application through the system to determine whether the decision changes once the error is corrected.

### *3.4. Counterfactual Explanations and Contestability: A Pessimistic (Interim) Conclusion*

Counterfactuals are often the preferred method for recourse recommendations, and the standard approach is to define a metric or procedure for selecting the most useful counterfactual for the decision subject. This approach, however, does not translate well to contestability.

Unlike recourse, contestability is adversarial: the interests of the decision-maker and the decision-subject diverge. This makes the Rashomon problem particularly acute. A well-defined procedure is needed to prevent decision-makers from selectively providing counterfactuals that lack information relevant for contesting a decision. Without such safeguards, contestability risks becoming an empty promise.

Specifying what must be provided is non-trivial. First, multiple counterfactual explanations may be necessary because contestability can rest on different grounds and involve different hypotheses about what went wrong. Second, personalization poses a challenge similar to that in recourse. In recourse, costs of changes vary across individuals—some can more easily find higher-paying jobs, others can pursue further education. It is therefore infeasible to use a single cost metric to automate the selection of counterfactual explanations most beneficial to all individuals. Likewise, in contestability, individuals' backgrounds shape the grounds on which they can contest and the cases they can build. A one-size-fits-all procedure for generating counterfactuals is unrealistic, yet greater personalization increases costs.

Moreover, contestability demands more domain and statistical knowledge than recourse. For recourse, domain knowledge concerns actionable life changes, which, while difficult, are easier to understand for a layperson. By contrast, recognizing a counterfactual as evidence for contesting a decision often requires more sophisticated reasoning and, in many cases, legal, domain or statistical expertise, as illustrated by Behm's and Yaojun's cases.

Focusing on contestability rather than recourse, the upshot, then, is that counterfactual explanations are not as user-friendly or cost efficient as we might have thought after all.

## 4. *Saving Contestability*

The preceding section paints a bleak picture regarding the way that counterfactual explanations enable contestability. Two possible responses are available: The first one is to join the ranks of critics of the right to explanation (Taylor, 2024; Grote & Paulo, 2025; Karlan & Kugelberg, 2025) and argue that the information required to enable contestability turns out to be intolerably costly to provide. The second one tries to save what can be saved by suggesting ways in which, with certain adjustments, counterfactual explanations can be made more fitting to serve the desired function. This section sets out to explore the second strategy. Our claim is not that there is a simple technical fix to overcome the various structural obstacles that counterfactual explanations face. Instead, it is about proposing ameliorative strategies.

Our positive contribution is centered around two pillars: (i) We advocate for a multi-shot approach to counterfactual explanations, and (ii) we discuss how background knowledge can be provided, so that decision-subjects can assess counterfactual explanations in a meaningful way.

### *4.1 Multi-Shot Approaches to Counterfactual Explanations*

Counterfactual explanations can, in some cases, provide evidence useful for contesting a decision. However, they face the Rashomon problem, and there is no procedure for generating a single counterfactual explanation that suffices for contestability. Moreover, to use counterfactual explanations meaningfully in contesting a decision, the user must already have some background knowledge about the kind of evidence they are seeking. The first challenge points to the need for a user-initiated, multi-shot approach, while the second suggests the involvement of someone in a supporting role, such as a case worker.

Multi-shot approaches (Mothilal et al., 2020; Russel, 2019; Wachter et al., 2017) as well as decision-subject-initiated interactive approaches (Wang et al., 2023; Citron & Pasquale, 2014; Hildebrandt, 2006) have both been previously proposed and studied in the context of recourse. The former emphasizes giving users access to a diverse set of explanations, while the latter focuses on allowing users to explore and shape the explanation most useful to them. Combining these ideas, a multi-shot, decision-subject-initiated approach positions users not as passive recipients of counterfactual explanations but as active participants who can query the model multiple times and tailor the explanations to their needs.

It is important that the counterfactual explanations are multi-shot because, again, there is no procedure to generate a single counterfactual explanation that fits the bill. Behm might suspect that he would have gotten the job if he had scored higher on the personality test, and also that he would have gotten it if he had been a different race. A multi-shot approach allows him to confirm or rule out multiple hypotheses, and user-initiation ensures that the explanations he obtains are ones that align with his concerns, rather than an unhelpful counterfactual such as *if you had had a PhD, you would have gotten the job*, which is true but irrelevant for contesting.

This multi-shot, decision-subject-initiated approach offers several advantages. First, it shifts power from decision-makers to decision-subjects. In doing so, it helps mitigate concerns about explanation hacking (Sullivan & Kasirzadeh, 2025), the risk that decision-makers might selectively provide explanations that serve their own interests rather than those of the decision-subjects.

Second, it addresses the personalization problem, or what Barocas et al. (2020) refer to as the 'autonomy paradox'. Each decision-subject has unique circumstances and is in the best position to understand them. In Behm's case, for example, details about his bipolar disorder and past job reports showing that his condition did not impair performance are known primarily to him, not to others. Decision-subject-initiated queries can leverage this personal knowledge, increasing the likelihood of uncovering evidence useful for contestation by allowing individuals to shape their queries based on their background knowledge.

There are, however, also problems with multi-shot decision-subject-initiated counterfactual explanations. First, allowing the users to query the model at will may pose serious privacy concerns (Berning et al., 2024). There is a risk that a model can be reconstructed, or sensitive information can be inferred from counterfactual information (Tramèr et al., 2016; Goethals et al., 2023). This risk can be mitigated, and there is already ongoing work on privacy-preserving counterfactual explanations (An & Cao, 2024; Vo et al. 2023; Yang et al., 2022; Patel et al., 2022). For the privacy-preserving mechanisms to work, however, the number of queries needs to be limited.[13] From a normative standpoint, this is not ideal, since the more queries a decision-subject can have, the more likely they are to obtain the information they need to contest. However, as is often the case for many normative issues surrounding machine learning, we need to navigate the trade-off between privacy and the effectiveness of counterfactual explanations by considering the stakes of the context.

The second challenge with multi-shot, decision-subject-initiated counterfactual explanations is the epistemic barrier in identifying relevant evidence.

Imagine going to a medical appointment because you have some unexplained headache. When the physician asks you whether there have been lifestyle changes that could explain this, you cannot think of anything that might be relevant. Of course, you are the one person in the world who knows yourself best. You have access to all the facts about your lifestyle, but the problem is that you lack self-reflexive knowledge about which aspects of your own situation are relevant to your headache.

The same problem exists in the context of counterfactual explanations and contestability. The decision-subject might know their own situation well, but fail to identify which aspects are relevant for contesting. This is presumably one of the key elements in Behm's case. He knows that he has bipolar disorder, but might not have realized that it was relevant for the personality test or the job that he was applying to.[14]

Even if we assume that decision-subjects know their own circumstances well, they may still lack the knowledge of which explanations to seek and how to use them effectively as evidence. Moreover, given the limited budget of queries, the decision-subject might not know how to use them in the most efficient way. While multiple queries increase the decision-subject's chances of finding relevant evidence, there is no guarantee they will succeed, or that such evidence even exists. This raises an interesting question: how should the limited budget of queries be allocated? One strategy is to test hypotheses in order of plausibility; another is to diversify queries across different dimensions (e.g., demographic, behavioral, financial) to maximize coverage.

One way to constrain the hypothesis space and help decision-subjects to find relevant counterfactuals could be to provide them with model cards (Mitchell et al., 2019). Although, standardly, they just report information about model types and performance metrics, we can think of possible extensions that would make them more fitting for the purpose of enabling contestability. For example, they might provide summary statistics about the most relevant features used by the model on average.

[13] For a technical explanation for why privacy preservation requires limited queries, see Dwork & Roth (2014) on privacy budget.

[14] Miller (2023) provides a more general account of the challenges that decision-subjects face to make sense of algorithmic explanations.

However, addressing the challenge of selecting and scrutinizing the *right* counterfactual explanations may ultimately require a more resource-intensive solution: providing case workers, as suggested in Vredenburgh (2022). In public assistance programs and social work, case workers help individuals navigate systems by assessing needs and connecting them to resources. Similarly, in this context, case workers would bring domain, legal, and statistical expertise to guide decision-subjects in formulating queries and making the most efficient use of counterfactual explanations to uncover evidence for contesting.

In the case of a mysterious headache, while the patient cannot identify which aspects of their lifestyle might explain this headache, a good physician knows how to ask the right questions to uncover relevant information and connect the dots that the laypeople themselves cannot connect. A case worker could play an analogous role to the physician and assist the decision-subject in identifying aspects of their lives that might be relevant for contestation. This alleviates the difficulty posed by the layperson's lack of self-reflexive knowledge. For example, after learning about Behm's mental health history and past job reports, a case worker could suggest querying the model on how his personality test affected the hiring decision.

The case worker's familiarity with the model, the domain, and previous cases can also increase the chances of successful inquiries. For example, a case worker with medical knowledge and experience from previously contested cases might advise Yaojun to query the model about how his lifestyle choices and medical conditions influence the predictions about him.

Our proposal to include case workers and provide user-initiated multi-shot explanations is primarily a pragmatic response to the challenges and constraints outlined in the preceding sections. We do not suggest that our approach is the only way that contestability ought to function, nor do we claim that legal or institutional procedures ought to be structured in this specific way; considerations such as cost, privacy, and efficiency may reasonably weigh against our proposed implementation. However, we emphasize that without additional structural and institutional support—of which our proposal is only one example—counterfactual explanations alone are unlikely to be sufficient. For this reason, we caution against treating counterfactual explanations as a low-cost, standalone solution.

### *4.2 Revisiting Costs*

As the need for detailed background knowledge increases, and the case workers assume a more prominent role in selecting and scrutinizing counterfactual explanations, the costs also rise in several forms. In particular, understanding of how a machine learning model arrives at its decisions enables decision-subjects to strategically manipulate certain attributes to achieve a desired outcome. Because such strategic behavior is more readily available to people who have traditionally occupied a privileged position in society, this can, in the long term, exacerbate structural inequalities (Hardt et al., 2016; Hu et al., 2019). Relatedly, the provision of information necessary for contestability, as well as case workers, can conflict with the interest in protecting business secrets or place demands of transparency onto institutions that arguably exceed current standards. This inevitably raises the worry about whether such a double-standard (Zerilli et al., 2019) can be reasonably justified.

Still, as forceful as critiques of the right to explanation may be, it is crucial to point out that their proposed solutions, too, come at a high price: If we shift the target to assessing the societal effects of machine learning models (Taylor, 2024), move toward a medical model, where the safety of machine learning models is established by independent institutions (Grote & Paulo, 2025), or outright deny the existence of the right to explanation (Karlan & Kugelberg, 2025), we take individuals' agency off the equation. In light of this, our amelioration strategy represents a modest attempt to preserve the individual's agency. Consequently, we leave the question of how much cost is tolerable for the right to explanation for future work.

## 5. *Conclusion*

The paper's starting point was the concern that the automation of consequential decisions through machine learning models across society poses a threat to our agency. Our main contribution, in turn, was to scrutinize the extent to which counterfactual explanations enable the backward-looking agential ability of contestability to be reinstated. For this purpose, we developed an account of contestability, and demarcated it from adjacent concepts in the discourse surrounding the right to explanation, such as justification and recourse.

However, upon analyzing whether counterfactual explanations are indeed capable of enabling contestability, the results were brittle: Often, the necessary information is too complex, too context-dependent or requires too much domain-knowledge to be meaningfully captured by a single counterfactual. As a remedy, we argued that with certain adjustments, counterfactual explanations can be made more adequate for their intended purpose. These adjustments center around a multi-shot approach that allows decision-subjects to query a model for counterfactuals a number of times, the provision of supplementary background information via model cards, and the support of case-workers. While this solution cannot resolve all concerns about the legitimacy of the right to explanation, it presents an empirically and technically tractable way forward that could inform future research and practices.

**Conflicts of Interest**:

The authors declare none.

**Acknowledgements:**

Alice C.W. Huang and Thomas Grote contributed equally to the paper. We thank two anonymous reviewers, the participants at the 6th Upstate Workshop on AI and Human Values at the University of Rochester and Kate Vredenburgh for helpful comments and discussions.

**Funding:**

Alice C.W. Huang acknowledges funding by the Social Science and Humanities Research Council, Grant Number: 430-2025-00038.